\documentclass[%
 aip,jap,
 amsmath,amssymb,
 reprint,%
]{revtex4-2} 

\usepackage{graphicx}
\usepackage{dcolumn}
\usepackage{bm}

\usepackage[utf8]{inputenc}
\usepackage[T1]{fontenc}
\usepackage{mathptmx}
\usepackage{etoolbox}
\usepackage{natbib}
\usepackage[colorlinks=true,linkcolor=blue,citecolor=blue,urlcolor=blue]{hyperref}

\renewcommand{\d}{{\rm d}}  
\newcommand{\e}{{\rm e}}  
\newcommand{\imai}{{\rm i}}

\makeatletter
\def\@email#1#2{%
 \endgroup
 \patchcmd{\titleblock@produce}
  {\frontmatter@RRAPformat}
  {\frontmatter@RRAPformat{\produce@RRAP{*#1\href{mailto:#2}{#2}}}\frontmatter@RRAPformat}
  {}{}
}%
\makeatother

\begin{document}

\preprint{Levels for Quantum Cascade Lasers}

\title[Levels for Quantum Cascade Lasers]{Orthonormal and periodic levels for Quantum Cascade Laser simulation}
\author{Zakaria Mohamed}
\author{D. Ekin {\"O}nder} 
\altaffiliation[Now at ]{2550 Engineering, G{\"o}teborg, Sweden}
\author{Andreas Wacker}
\email{Andreas.Wacker@teorfys.lu.se}
\homepage{https://www.matfys.lu.se/staff/faculty/andreas-wacker/}
\affiliation{Mathematical Physics and NanoLund, Lund University, Box 118, 22100 Lund, Sweden}

\date{\today, accepted by Journal of Applied Physics}

\begin{abstract}
A Python package to evaluate Wannier, Wannier-Stark, and EZ (both Energy and location Z resolved) levels for Quantum Cascade Lasers is presented. We provide the underlying theory in detail with a focus on the orthonormality and periodicity of the generated states. 
\end{abstract}

\maketitle

\section{Introduction}
The operation of Quantum Cascade Lasers
(QCLs)\cite{FaistScience1994,FaistBook2013,BotezBook2023} is based on
optical transitions between quantized levels in the conduction band of
a semiconductor heterostructure. Further such levels are employed for
an effective population of the upper laser level and the emptying of
the lower laser level, which is facilitated by a sequence of tunneling and scattering
processes at the designated bias. In order to provide light
amplification over a larger spatial range, a carefully designed
sequence of layers, referred to as module in the following, 
is repeated periodically about
30-300 times.  Each of these modules contains a laser transition, so
that the electron flow resembles a cascade.

The design of QCLs requires reliable simulation schemes, see
Refs.~\onlinecite{JirauschekApplPhysRev2014,WackerBook2023} for an overview. In a first
step this needs to establish the energy levels for a given sequence of
semiconductor layers. Here the periodic repetition of the module is
typically reflected by the requirement that the levels in a given
module are periodically repeated in the neighboring modules
(\textit{periodicity criterion}). Alternatively, few modules embedded between contacts can be simulated\cite{KubisPRB2009}.
Furthermore, the \textit{orthonormality of the levels} is highly relevant, as this is a common requirement for basis states and often
tacitly assumed in concepts such as Fermi's golden rule.

The periodicity criterion and orthonormality are actually difficult to
achieve simultaneously. The reason is that common approaches require
boundary conditions to avoid escape to infinity. In this case, levels
located in different modules do not satisfy the periodicity criterion,
as their distance to the boundary differs. Alternatively, a central
module is identified and levels in neighboring modules are defined by
shifting the levels from the central module. In this case, levels in
neighboring modules are solutions of a setup with shifted boundaries
and thus intermodule orthogonality is not guaranteed either. Albeit these
orthogonality violations may be small\cite{BurnettIEEETransTHzSciTechn2018} they
constitute a matter of concern.

A second issue affecting the orthogonality of levels is the common
use of energy-dependent effective masses, see,
e.g., Refs.~\onlinecite{KubisPRB2009,BaranovIEEEJSelTopQuantEl2015,KatoJAP2019}.
These are particular important for infrared QCLs with level energies
of several 100 meV above the conduction band edge where the conduction
band becomes non-parabolic, see e.g. Ref.~\onlinecite{KoteraJAP2013}
and references cited therein. If the Hamiltonian contains
energy-dependent components, the solutions obtained at different energies
are not orthogonal. 

Over the last decades our group developed simulation techniques for QCLs,
where we strived to overcome the issues mentioned
above. From the very start we used a basis of Wannier
states\cite{LeePRB2002} to combine the periodicity condition with
orthogonality. Later, we implemented non-parabolicity
\cite{LindskogSPIE2013} by applying a two-band model
\cite{WhitePRL1981,BastardPRB1981,SirtoriPRB1994} explicitly. In this article we give
a comprehensive discussion of the underlying concepts. We also
carefully overhauled our codes and provide a new Python-based
simulation package \texttt{resource\_QCL} to generate sets of orthonormal and
periodic levels for a variety of QCL simulations. The package is provided in the supplemental material and can be used under a CC-BY license.

\section{Generating Wannier states}
\begin{figure}
\includegraphics[width=\columnwidth]{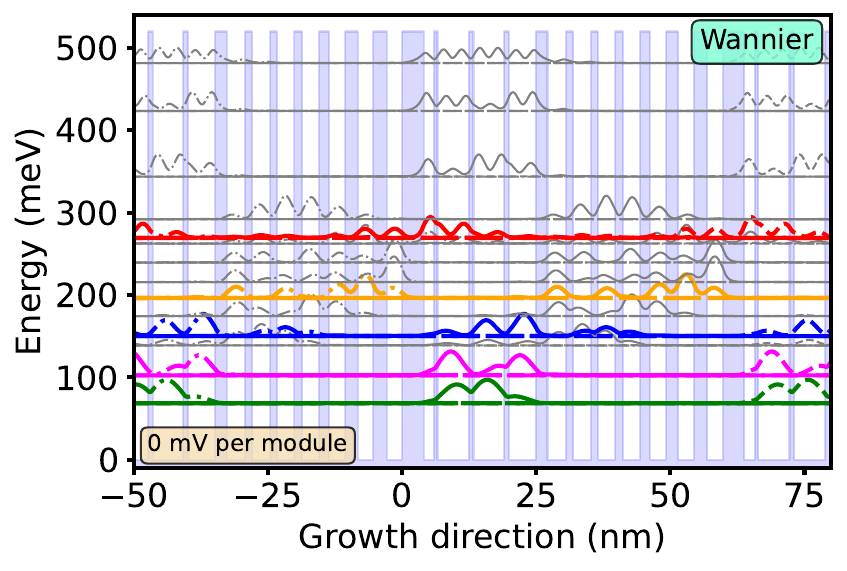}
\caption{\label{fig:BeckWannier0} Wannier states for the QCL of Ref.~\onlinecite{BeckScience2002}. The band gap is depicted by the grey/blue area. The lines show the probability density $w^{\nu,n}_c(z)^2+w^{\nu,n}_v(z)^2$ for the 14 lowest Wannier states (label $\nu$), where the baseline is shifted by the respective level energy. Thick colored lines indicate levels related to (from top to bottom) the upper laser, the injector, the lower laser, and the two extraction levels. Full lines refer to levels in the central module ($n=0$) ranging from 0 to 59.8 nm. Dashed and dot-dashed lines are periodic repetitions to the right and left module ($n=\pm 1$), respectively.  }
\end{figure}

To avoid boundary conditions, which are incompatible with the periodicity condition, we start by calculating Bloch functions for an infinite periodic repetition of the modules without bias. These are mapped on Wannier functions $w^{\nu n}$ by a unitary transformation. Here, $n$ specifies the module to which they are localized and $\nu$ distinguishes different levels in a module.
Most importantly, they satisfy the periodicity criterion and constitute a basis, which is complete within the energy range covered by the Bloch bands considered.
In particular, we obtain an explicit matrix representation \eqref{eq:H_het_Wannier} of the Hamiltonian for the heterostructure. Fig.~\ref{fig:BeckWannier0} shows an example for these Wannier levels. Here, we consider the sample of Ref.~\onlinecite{BeckScience2002}, which achieved room-temperature CW operation by double-phonon extraction using two extraction levels with a spacing of the optical phonon energy.

Nonparabolicity is treated by a two-band model and keeping both the conduction and valence band components $(w^{\nu n}_c(z),w^{\nu n}_v(z))^\textrm{tr}$. Thus orthonormality is ensured, see Eq.~\eqref{eq:Wannier_Orthonorm} below. This task is performed by the routine \texttt{resource\_QCL.calcWannier1} which establishes the relevant data based on the following physical input parameters:
\begin{description}
\item[\texttt{thlist}] list of layer thicknesses $w_i$ in one module
\item[\texttt{Eclist}] list of conduction band offsets $V_i$ for the layers
\item[\texttt{mclist}] list of effective masses  $m_c^*$ at the $\Gamma$ point
\item[\texttt{E\_Kane}] the common Kane energy $E_\textrm{Kane}$
\end{description}
In this section the underlying theory and numerical issues are detailed. 
\subsection{Two-band formalism}
The two-band model  considers states
$(\Psi_c(z),\Psi_v(z))^\textrm{tr}$, which contain a conduction and
valence band contribution in order to include non-parabolicity in the
conduction band for planar systems. This model is applicable only along a single spatial direction, typically the growth direction denoted as $z$. In the absence of an external potential, the
two-band Hamiltonian is given by the matrix \cite{WhitePRL1981}
\begin{equation}
\hat{H}^\textrm{het}=
\begin{pmatrix} E_c(z) 
    & \frac{p_{cv}}{m_e}\frac{\hbar}{\imai}\frac{\partial}{\partial
    z}\\ \frac{p_{vc}}{m_e}\frac{\hbar}{\imai}\frac{\partial}{\partial
    z}  & E_v(z)
\end{pmatrix} \, .
\label{Eq2BandHam}
\end{equation}
$E_c(z)$ is the conduction-band offset, which
depends on the specific material at position $z$. Correspondingly,
$E_v(z)$ mimics the valence band offset. The stationary Schr{\"o}dinger equation reads
\begin{equation}\hat{H}^\textrm{het}
\begin{pmatrix}\Psi_c(z)\\ \Psi_v(z)\end{pmatrix}
=E\begin{pmatrix}\Psi_c(z)\\ \Psi_v(z)\end{pmatrix}
\label{eq:Schrodinger}
\end{equation}
where the lower component  provides
\begin{equation}
  \Psi_v(z)=\frac{1}{E-E_v(z)}\frac{p_{vc}}{m_e}\frac{\hbar}{\imai}
  \frac{\partial \Psi_c(z)}{\partial z}\, .
    \label{EqStatSE_2band_v}
\end{equation}
Inserting into the upper component gives the Schr{\"o}dinger-like equation
\begin{equation}
\left[E_c(z)-\frac{\partial}{\partial
    z}\frac{\hbar^2}{2m_c(E,z)} \frac{\partial}{\partial
    z}\right] \Psi_c(z)=E \Psi_c(z)\label{EqStatSE_2band_c}
\end{equation}
with the energy-dependent mass
\begin{equation}
m_c(E,z)=\frac{m_e^2[E-E_v(z)]}{2 |p_{cv}|^2}\, ,
\label{EqStatSE_effmass}
\end{equation}
which also
depends on the material at position $z$.
Identifying the known effective mass $m^*_c$ at the $\Gamma$ point as
$m_c(E_c)$, we find the relation
\[
m^*_c(z)=\frac{m_e^2(E_c(z)-E_v(z))}{2 |p_{cv}|^2}
\]
so that the values of $E_c(z),E_v(z),p_{cv}$ and $m^*_c(z)$ are
intrinsically related to each other within the two-band model. The
value of $p_{cv}$ is required to be the same for all materials
involved in a given heterostructure within the two band model. Its value is provided by the Kane
energy\cite{SirtoriPRB1994} $E_\textrm{Kane}=2|p_{vc}|^2/m_e$, which
is used as an input parameter. Focusing on the conduction band, we use literature values for $m^*_c(z)$ and $E_c(z)$ for the respective material at $z$ as further input parameters. Hence we apply 
\begin{equation}
E_v(z)=E_c(z)-2|p_{vc}|^2\frac{m^*_c(z)}{m_e^2}=E_c(z)-E_\textrm{Kane}\frac{m^*_c(z)}{m_e}\, ,
\label{EqCalcEv}
\end{equation}
for the effective valence band offset.
This approximation deviates from the actual material value, as expected within the framework of the two-band model, which simplifies the intricate multi-band structure of the valence band. Here, its sole purpose is to quantify the energy-dependent effective mass $m_c(E,z)$ from Eq.~\eqref{EqStatSE_effmass}. 
Finally, the phase
of the momentum matrix elements is chosen as
\[
p_{vc}=\imai |p_{vc}|=\imai\sqrt{ \frac{m_eE_\textrm{Kane}}{2}}
\textrm{ and consequently }p_{cv}=-\imai |p_{vc}|\, ,
\]
which provides a real Hamiltonian \eqref{Eq2BandHam} and consequently
the option to establish real Wannier functions below.

Note, that the solutions $\Psi^n_c(z)$ of the effective Schr{\"o}dinger equation \eqref{EqStatSE_2band_c} at different eigenenergies $E_n$ are not necessarily orthogonal due to the energy-dependent mass. However, adding the valence band component from Eq.~\eqref{EqStatSE_2band_v} provides after normalization
\[
\int \d z \, \left[\Psi^{m*}_c(z)\Psi^n_c(z)+\Psi^{m*}_v(z)\Psi^n_v(z)\right]=\delta_{mn}
\]
as they are actually solutions of the energy-independent Hamiltonian 
\eqref{Eq2BandHam}. The main concept in the following is to start with othonormal Wannier functions, see Eq.~\eqref{eq:Wannier_Orthonorm}, and inherit this property to other sets by unitary transformations.

\subsection{Bloch States}
Now we consider a QCL, where each module consists of a sequence of $N$
layers with thicknesses $w_0,w_1,\ldots w_{N-1}$  extending from $z_i$
to $z_{i+1}=z_i+w_i$. Here we set $z_0=0$ and denote $z_N=d$. As the
modules are periodically repeated, we have a periodic structure with
period $d=\sum_{i=0}^{N-1}w_i$. Consequently, for vanishing external 
potential, the eigenstates are Bloch states
$(\Psi^{q\nu}_c(z),\Psi^{q\nu}_v(z))^\textrm{tr}$ as characterized by
the Bloch vector $q$ (with $-\pi/d< q \le \pi/d$ covering the one-dimensional Brillouin zone) and the band index $\nu$. The Bloch condition reads
\begin{equation}
\begin{pmatrix}\Psi^{q\nu}_c(z+d)\\ \Psi^{q\nu}_v(z+d)\end{pmatrix}=\e^{\imai qd}
\begin{pmatrix}\Psi^{q\nu}_c(z)\\ \Psi^{q\nu}_v(z)\end{pmatrix}
\label{EqBlochCond_2}
\end{equation}
for all $z$. In order to calculate the Bloch states we note that in a
region $z_{i}<z<z_{i+1}$ of constant $E_c(z)=V_i$ and constant
$m_c(E,z)=m_i(E)$, Eq.~\eqref{EqStatSE_2band_c} is solved by
\begin{equation}\begin{split}
    &\Psi^i_c(z)=C_i\cos\left[k_i  \left(z-z_i-\frac{w_i}{2}\right)\right]
    +\frac{D_i}{k_i} \sin\left[k_i  \left(z-z_i-\frac{w_i}{2}\right)\right]\\
    &\textrm{with } k_i=\frac{\sqrt{2m_i(E)(E-V_i)}}{\hbar} \quad\textrm{for }z_i<z<z_{i+1}
\label{EqSolutionRegion_1}
\end{split} \end{equation}
This provides a finite nonvanishing result for the $D_i$-term at $k_i=0$, where we implement
$\frac{1}{k_i} \sin(k_i z)= z\;\textrm{sinc}(k_iz/\pi)$ with the normalized sinc function $\textrm{sinc}(x)=\sin(\pi x)/(\pi x)$ as defined in the NumPy library of Python.

For $E<V_i$, $k_i=\imai \lambda_i$ is purely imaginary and we get the equivalent
representation
\[\begin{split}  
 &\Psi^i_c(z)=C_i \cosh \left[\lambda_i   \left(z-z_i-\frac{w_i}{2}\right)\right]+\frac{D_i}{\lambda_i}\sinh\left[\lambda_i   \left(z-z_i-\frac{w_i}{2}\right)\right]\\
 &\textrm{with }\lambda_i =\frac{\sqrt{2m_i(E)(V_i-E)}}{\hbar}
\end{split}\]
Our choice for the middle point of the layers $z_i+\frac{w_i}{2}$
as reference position limits the exponential growth of the hyperbolic functions,
which reduces numerical instabilities.

At the boundary $z_{n+1}$ between two layers, the functions $\Psi_c(z)$
and $\Psi_v(z)$ should be continuous. From
Eq.~\eqref{EqStatSE_2band_v} we find, that this is equivalent to the two equations
\[\begin{split}
\Psi^n_c(z_{n+1}-0^+)&=\Psi^{n+1}_c(z_{n+1}+0^+)\\
\frac{1}{m_n}\frac{\partial \Psi^n_c(z)}{\partial
  z}_{|z=z_{n+1}-0^+}&=\frac{1}{m_{n+1}}\frac{\partial
  \Psi^{n+1}_c(z)}{\partial z}_{|z=z_{n+1}+0^+}\, ,
\end{split}\]
where $0^+$ stands for an infinitesimal positive quantity (more
formally to be replaced by an $\epsilon>0$ with $\lim_{\epsilon\to
  0}$). Inserting Eq.~\eqref{EqSolutionRegion_1} provides the
relation
\begin{widetext}\[\begin{split}
&\begin{pmatrix}
  C_{n+1}\\ D_{n+1}
  \end{pmatrix}=\tilde{\cal M}_{n}\begin{pmatrix}
  C_{n}\\ D_{n}
\end{pmatrix}
\quad\textrm{with } \tilde{\cal M}_{n}=
\begin{pmatrix}
c_{n} c_{n+1}- \beta_n k_n^2 s_{n} s_{n+1} & s_{n}c_{n+1}+ \beta_n  c_{n}s_{n+1} \\
  -c_{n}s_{n+1}k_{n+1}^2 - \beta_n k_n^2 s_{n}c_{n+1} & -k_{n+1}^2s_{n}s_{n+1}+ \beta_n c_{n}c_{n+1}
\end{pmatrix}\\
&\textrm{where}\quad 
\beta_n=\frac{m_{n+1}}{m_{n}}\qquad c_n=
\cos\left(\frac{k_nw_n}{2}\right)
\qquad s_n=\frac{1}{k_n}\sin\left(\frac{k_nw_n}{2}\right)=\frac{w_n}{2}\textrm{sinc}\left(\frac{k_nw_n}{2\pi}\right)
\end{split}\]\end{widetext}
Here the values $c_n$ and $s_n$ are real both for real and (purely) imaginary $k_n$, so that the entire matrix is real.
Note that the  matrices $\tilde{\cal M}_n$ depend on $E$ via $k_n$
and $m_n$.  Applying the Bloch condition \eqref{EqBlochCond_2} for the
conduction band component\footnote{The valence band component
  $\Psi^{q\nu}_v(z)$ satisfies the same condition due to
  Eq.~\eqref{EqStatSE_2band_v}.}  $\Psi^{q\nu}_c(z+d)=\e^{\imai qd}
\Psi^{q\nu}_c(z)$ for  $z_0\le z<z_1$, where $z_N\le z+d<z_{N+1}$ is
in region $N$, which has the same properties as region 0
(i.e. $m_{N}=m_0$ and $V_{N}=V_0$) requires
\begin{equation}
\e^{\imai qd}\begin{pmatrix} C_{0}\\ D_{0}
\end{pmatrix}
=
\begin{pmatrix}
  C_{N}\\ D_{N}
\end{pmatrix}=
\tilde{\cal M}
\begin{pmatrix}
  C_{0}\\ D_{0}
\end{pmatrix}
\label{eq:eigenvalue}
\end{equation}

with $\tilde{\cal M}=\tilde{\cal M}_{N-1} \tilde{\cal M}_{N-2}
\ldots \tilde{\cal M}_{1} \tilde{\cal M}_{0}$. 
This only allows for non-vanishing $C_0,D_0$ if $\textrm{det}\{\tilde{\cal M}
-\e^{\imai qd}{\cal I}\}=0$. As $\tilde{\cal M}$ depends on energy this
provides a set of energies $E_{\nu}(q)$ with $\nu=1,2,\ldots$ for each
value of $q$. (Restricting to conduction band states, we consider
only solutions, where $E$ is larger than the minimal value of the set
$\{V_n\}$.)

Straightforward evaluation provides det$\{\tilde{\cal M}_n\}=\beta_n$. Thus
det$\{\tilde{\cal M}\}=m_N/m_0=1$ due to the periodicity. This provides
\[\begin{split}
0=&\textrm{det}\{\tilde{\cal M}-\e^{\imai qd}{\cal I}\}
= 1-(M_{00}+M_{11})\e^{\imai qd}+\e^{2\imai qd}\\
=&
\e^{\imai qd}\left[2\cos(qd)-(M_{00}+M_{11})\right]
\\
\Leftrightarrow & M_{00}+M_{11}=2\cos(qd)
\end{split}\]

For the specific energies $E_{\nu}(q)$, a solutions of Eq.~\eqref{eq:eigenvalue} is given by \begin{equation}
\begin{pmatrix}C_0\\D_0
\end{pmatrix}= \begin{pmatrix}\tilde{M}_{11}-\e^{\imai qd}\\-\tilde{M}_{10}
\end{pmatrix}
+0.5 \begin{pmatrix}-\tilde{M}_{01}\\\tilde{M}_{00}-\e^{\imai qd}
\end{pmatrix}
\label{EqEigenvector}
\end{equation}
where we added solutions for the first and second row in order to avoid problems with accidental small values.  Furthermore, we directly get all coefficients
\[
\begin{pmatrix}
  C_{n}\\ D_{n}
\end{pmatrix}=\tilde{\cal M}_{n-1}\tilde{\cal M}_{n-2}\ldots \tilde{\cal M}_{0}
\begin{pmatrix}
  C_{0}\\ D_{0}
\end{pmatrix}
\]
which allows us to construct the wave function $\Psi^n_c(z)$ for
arbitrary $z$ in the interval $0<z\le d$ by evaluation of
Eq.~\eqref{EqSolutionRegion_1} in the appropriate region. [Outside
  this central module, the periodicity condition \eqref{EqBlochCond_2}
  is applied to evaluate  $\Psi^n_c(z)$ based on the result in the
  central module.]  Furthermore, $\Psi^n_v(z)$ is obtained from
Eq.~\eqref{EqStatSE_2band_v}. By rescaling $C_0^\nu(q)$ and $D_0^\nu(q)$ the
normalization
\begin{equation}
\int_0^d\d z \left[ |\Psi^{\nu q}_c(z)|^2+|\Psi^{\nu
    q}_v(z)|^2\right]=1
\label{EqNormBloch}
\end{equation}
is achieved. This procedure defines the initial Bloch functions. 
As the Hamiltonian is real we have $E_\nu(-q)=E_\nu(q)$ and our choice \eqref{EqEigenvector} guarantees  $C_n(-q)=C_n(q)^*$ and $D_n(-q)=D_n(q)^*$ so that
\begin{equation}
\begin{pmatrix}\Psi^{q\nu}_c(z) \\ \Psi^{q\nu}_v(z)\end{pmatrix}^*=
\begin{pmatrix}\Psi^{-q\nu}_c(z)\\ \Psi^{-q\nu}_v(z)\end{pmatrix}
\label{EqBlochPhaseMinusq}
\end{equation}
Furthermore, the chosen Bloch funtions are periodic in $q$ and
continuous at $q=\pm\pi/d$.

In practice, we calculate the Bloch functions for a finite number
$(N_q)$ of $q$-values 
\begin{equation}
q_j=-\frac{\pi}{d}+\frac{\pi}{N_q d}+\frac{2\pi}{N_q d}j
\quad\textrm{for } j=0,1,2,\ldots N_q-1 \, ,
\label{eq:qvalues}
\end{equation}
which are chosen such that they cover the entire Brillouin zone with equal
spacing. Furthermore, for each $j$, we have a value $k=N_q-1-j$, so that 
both $q_j$ and $q_{k}=-q_j$ are in the set of chosen $q$-values.
The choice of $q_j$ values restricts to the eigenstates of a system of
length $L=N_qd$ with boundary conditions $\Psi(x+L)=(-1)^{(N_q+1)}\Psi(x)$.
  (For odd $N_q$ this is the standard Born von-Karman boundary condition). Thus, the eigenstates are orthogonal. Based on normalization for  a single
  period before, we thus find
\begin{equation}
\int_{-L/2}^{L/2}\d z\, 
\begin{pmatrix}\Psi^{q_j\nu}_c(z) &\Psi^{q_j\nu}_v(z)\end{pmatrix}^*
\begin{pmatrix}\Psi^{q_k\mu}_c(z)\\ \Psi^{q_k\mu}_v(z)\end{pmatrix}=
N_q\delta_{\nu\mu}\delta_{jk}\label{EqOrthonormBloch}
\end{equation}

As the Bloch states are eigenstates of the Hamiltonian $\hat{H}^\textrm{het}$ for the ideal layer sequence \eqref{Eq2BandHam}, we can write its matrix elements of
\begin{multline}
 \int_{-L/2}^{L/2}\d z\, 
\begin{pmatrix}\Psi^{q_j\nu}_c(z) &\Psi^{q_j\nu}_v(z)\end{pmatrix}^*\hat{H}^\textrm{het}
\begin{pmatrix}\Psi^{q_k\mu}_c(z)\\ \Psi^{q_k\mu}_v(z)\end{pmatrix}\\
=
E_\nu(q)N_q\delta_{\nu\mu}\delta_{jk}
\label{eq:HamBloch}
\end{multline}
In the continuum limit $L=N_qd\to\infty$, one needs to replace $N_q\delta_{jk}\to 2\pi/d\; \delta(q_j-q_k)$ at the right-hand side of Eqs.~(\ref{EqOrthonormBloch},\ref{eq:HamBloch}).

\subsection{Wannier States}
The Wannier functions are given by
\begin{equation}\begin{split}
\begin{pmatrix}w_c^{\nu,n}(z)\\w_v^{\nu,n}(z)\end{pmatrix}
=&\frac{d}{2\pi}\int_{-\pi/d}^{\pi/d}\d q\,  \e^{-\imai qnd}
\begin{pmatrix}\Psi^{q\nu}_c(z)\\ \Psi^{q\nu}_v(z)\end{pmatrix}\\
\approx&
\frac{1}{N_q}\sum_q \e^{-\imai q nd}
\begin{pmatrix}\Psi^{q\nu}_c(z)\\ \Psi^{q\nu}_v(z)\end{pmatrix}
\label{EqDefWannier}
\end{split}\end{equation}
and satisfy $w^{\nu,n}(z)=w^{\nu,0}(z-nd)$, so that it is sufficient to
calculate the $n=0$ terms. Here the $q$ sum runs over the values specified in Eq.~\eqref{eq:qvalues}. For the normalization \eqref{EqNormBloch} they satisfy
\begin{equation}
\int\d z
\begin{pmatrix}w_c^{\nu,n}(z) & w_v^{\nu,n}(z)\end{pmatrix}
\begin{pmatrix}w_c^{\mu,m}(z)\\w_v^{\mu,m}(z)\end{pmatrix}=\delta_{nm}\delta_{\nu\mu}
\label{eq:Wannier_Orthonorm}
\end{equation}
as given in Ref.~\onlinecite{BrunoPRB2007}. [This also  holds exactly for
  the approximation with a finite set of $q_j$ due to
  Eq.~\eqref{EqOrthonormBloch} where the integral is limited to the range $-L/2<z<L/2$.] Note that the Wannier functions are
real due to Eq.~\eqref{EqBlochPhaseMinusq}, which actually requires
the combined presence of $q_j$ and $-q_j$ in the sum for a finite number of
values.

Based on Eq.~\eqref{eq:HamBloch} we can express the Hamiltonian in the basis of the Wannier states  as
\begin{equation}
H^\textrm{het}_{\nu n,\mu m}=
E_{\nu |m-n|}\delta_{\nu\mu}
\textrm{ with }E_{\nu h}=
\frac{1}{N_q}\sum_q
E_\nu(q)\cos(hqd)\label{eq:H_het_Wannier}
\end{equation}
This provides the energies $E_{\nu 0}$ of the Wannier levels, as shown in Fig.~\ref{fig:BeckWannier0}, which are just the average energy of the respective Bloch band. 
The interperiod couplings  $E_{\nu 1}$,  $E_{\nu 2}, \ldots $  reflect the width and further details in the dispersion of the Bloch band. In most cases, one can restrict to $E_{\nu 1}$ and possibly $E_{\nu 2}$. (Exceptions exist for particular cases where the gaps between minibands vanish for energies above the barriers \cite{SirtoriAPL1994}.)

The Wannier functions evaluated by Eq.~\eqref{EqDefWannier} depend on
the phases chosen for the Bloch functions, which can be modified by an arbitrary
q-dependent phase factor $\e^{\imai \varphi_q}$ with the restriction
$\varphi_{-q}= -\varphi_{q}$. W.~Kohn demonstrated already in 1959
\cite{KohnPR1959} that choosing the
Wannier functions real at symmetry points provides often the best
localization of the Wannier functions. While superlattices have such
characteristic points with inversion symmetry, things are more
complicated for QCLs, where such a symmetry is lacking. However,
choosing all Bloch states real and positive at certain points is often
still a good choice, resulting in Wannier functions localized close to
the chosen point.

Alternatively, the more intricate procedure of Ref.~\onlinecite{BrunoPRB2007}, which we follow here, guarantees a minimal variance (MinVar) in $z$ with the probability distribution given by the square of the Wannier functions for each band. It requires that the
initial Bloch functions are continuous
and periodic with respect to the variable $q$ with period $2\pi/d$, as guaranteed by our choice \eqref{EqEigenvector}.
Then we calculate Eq.~(12, 13) of
Ref.~\onlinecite{BrunoPRB2007}
\[\begin{split}
 X_\nu(q)=&\imai \int_0^d\d z\, \e^{\imai
  qz}\begin{pmatrix}\Psi^{q\nu}_c(z)& \Psi^{q\nu}_v(z)\end{pmatrix}^*
\frac{\partial }{\partial q}\left[ \e^{-\imai qz}
\begin{pmatrix}\Psi^{q\nu}_c(z)\\ \Psi^{q\nu}_v(z)\end{pmatrix}\right]\\
\textrm{and }&x_\nu=\frac{d}{2\pi}\int_{-\pi/d}^{\pi/d}\d q\,  X_\nu(q)
\end{split}\]
and Eq.~(20) of Ref.~\onlinecite{BrunoPRB2007} determines the phases
\[
\varphi_\nu^\textrm{MinVar}(q)=\int_{0}^{q}\d q' [X_\nu(q')-x_\nu]
\]
Finally, we multiply our initial Bloch functions with index $\nu,q$ by the factor
$\e^{\imai \varphi_\nu^\textrm{MinVar}(q)}$ and use these in
Eq.~\eqref{EqDefWannier} to obtain Wannier
functions with minimal variance.

This procedure turns out to be robust for Wannier states derived from Bloch bands with energies below 
the conduction band edge of the barrier. For states with higher energy, large numbers of $N_q$ are required and orthonormality becomes less precise.

\section{States at finite bias}
\begin{figure}
\includegraphics[width=\columnwidth]{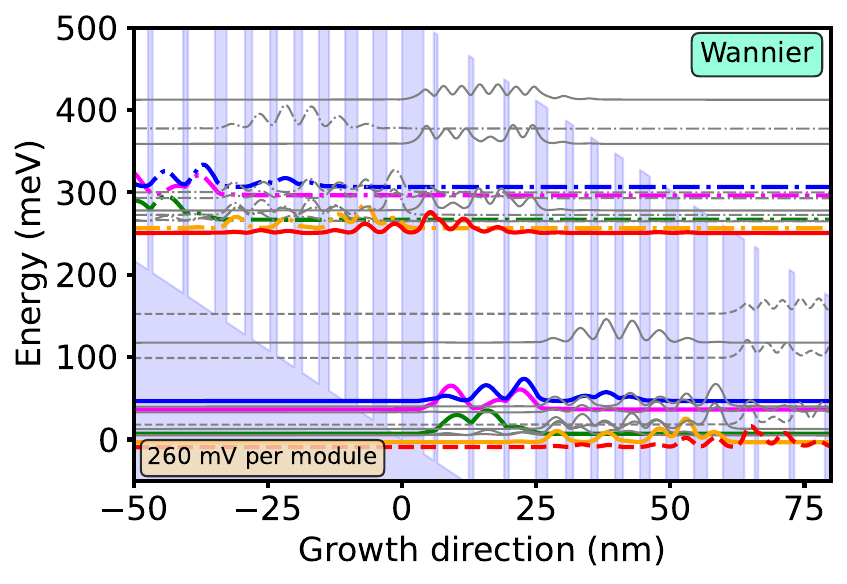}
\caption{\label{fig:BeckWannier260} Wannier levels as in Fig.~\ref{fig:BeckWannier0} under a constant electric field corresponding to a bias drop of 260 mV per module.}
\end{figure}

Under operation, the applied bias provides an additional scalar potential $\Phi(z)$ with the two-band Hamiltonian 
\[
\hat{H}^\Phi=-e\Phi(z)  \begin{pmatrix} 1 & 0\\0 & 1\end{pmatrix}
\, .
\] 
This tilts the energy landscape, as can be seen in Fig.~\ref{fig:BeckWannier260}. The matrix elements in the Wannier basis read:
\begin{equation}
H^\Phi_{\nu n,\mu m}=-e\int\d z\, \Phi(z)( w^{\nu n}_c(z) w^{\mu m}_c(z) +
w^{\nu n}_v(z) w^{\mu m}_v(z) )\label{eq:H_pot_Wannier}
\end{equation}
The diagonal elements shift the energies of the Wannier levels as illustrated in Fig.~\ref{fig:BeckWannier260}. This manifests their denotation (such as upper and lower laser levels) as provided in the caption of Fig.~\ref{fig:BeckWannier0}.

Here, we assume a constant electric field $F$ providing a linear potential $\Phi(z)=Fz$. ($F$ is counted positive, if electrons with charge $-e$ are accelerated towards the right). The mean-field from the electron distribution and the ionized dopands provides a further contribution $\Phi^\textrm{MF}(z)$ to $\Phi(z) $, which is periodically repeated for the modules and can be directly included in the package. However, its implementation requires to solve a kinetic equation for electron populations separately. Most importantly, we assume a constant bias drop $Fd$ per module, which is the main parameter in the following. This excludes the formation of field domains\cite{LuPRB2006,DharSciRep2014,AlmqvistEurPhysJB2019}.

\subsection{Wannier-Stark States}
We have an explicit expression for the matrix elements of the total Hamiltonian $\hat{H}^{F}=\hat{H}^\textrm{het}+\hat{H}^\Phi$ in the basis of Wannier levels, see Eqs.~(\ref{eq:H_het_Wannier},\ref{eq:H_pot_Wannier}). This Hamiltonian can be diagonalized and the corresponding energy eigenstates are related to the Wannier levels by a unitary transformation. These levels are orthonormal by construction.

The matrix elements of $\hat{H}^{F}$ in Wannier basis have the symmetry
\[
H^{F}_{\nu n+h,\mu m+h}=H^{F}_{\nu n,\mu m}- heFd \delta_{n,m}\delta_{\nu,\mu}
\]
This implies the labeling of the eigenlevels as 
$\psi^{\alpha n}(z)=\psi^{\alpha 0}(z-nd)$ and $E^\textrm{WS}_{\alpha n+h}=E^\textrm{WS}_{\alpha 0}-neFd$ for the infinite structure. These are known as \textit{Wannier-Stark} (WS) levels \cite{WannierPR1960}.
In our code, we repeat the central module \texttt{Nper} times in each direction and
choose the levels $\psi^{\alpha 0}(z)$ where the expectation value of $z$ is located within the central module. For other values of $n$, we apply the periodicity condition above for the functions and energies. We check explicitly the orthonormality between neighboring modules (default accuracy $10^{-4}$ in \texttt{resource\_QCL.calcZmat}). Otherwise, \texttt{Nper} should be increased. \texttt{Nper}$=3$ is sufficient under common operation conditions for most QCLs we studied. Larger values are in particular required if $eFd$ is small with respect to the range of Wannier energies $E_{\nu0}$. Thus, the WS states become problematic for low bias, where they become more and more extended, approaching the Bloch states for $eFd\to 0$. Similar to the Wannier and Bloch levels, the WS functions $\psi(z)$ have a valence and conduction band component.

\begin{figure}
\includegraphics[width=\columnwidth]{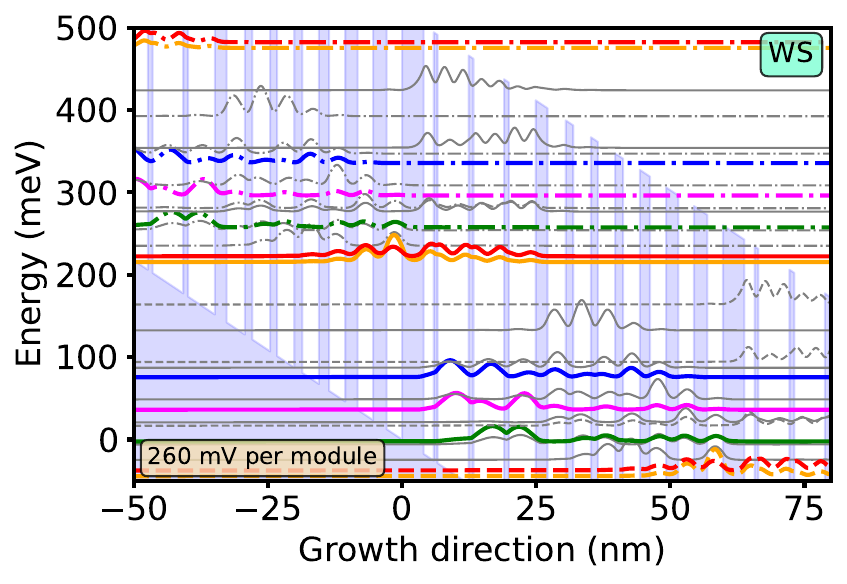}
\caption{\label{fig:BeckWS260} Wannier-Stark (WS) levels obtained by diagonalising the matrix of the Hamiltonian in the basis of Wannier levels from Fig.~\ref{fig:BeckWannier260}. See Fig.~\ref{fig:BeckWannier0} for labeling.}
\end{figure}

Fig.~\ref{fig:BeckWS260} shows the result. Comparing with the Wannier levels at the same bias drop per period in Fig.~\ref{fig:BeckWannier260}, the WS levels get typically repelled from each other resembling avoided crossings. The upper and lower extraction level are now approximately one and two optical phonon energies below the lower laser level, respectively, as targeted by the design.

Now we want to point out, how this cumbersome approach to establish energy eigenstates  via Bloch and Wannier functions is able to solve the problem of combined orthonormality and periodicity addressed in the introduction. The key problem in conventional calculations under applied bias  is the need for a spatial boundary. Otherwise, the wave functions leak out to infinity. 
Here we work in the basis of Wannier functions, where we applied a cut-off in energy by restricting to a finite number of Bloch bands. This cut-off is relative to the conduction band edge and thus tilted under applied bias. Therefore, the states cannot leak out to infinity as this would require the coupling to states at high energy above the conduction band (which would be incorrect within the two-band model anyway).

\subsection{EZ states}
If WS levels are close in energy, they sometimes appear to be linear combinations of more localized states. Examples can be seen around 210 meV and -5 meV in Fig.~\ref{fig:BeckWS260}. Such pairs of levels have to be treated with care, as they require a concise treatment of nondiagonal elements in the density matrix \cite{CallebautJAP2005}. Thus it is helpful to disentangle theses states into a pair of localized states. This, however, provides a nondiagonal Hamiltonian containing  tunneling matrix element between these levels. 

\begin{figure}
\includegraphics[width=\columnwidth]{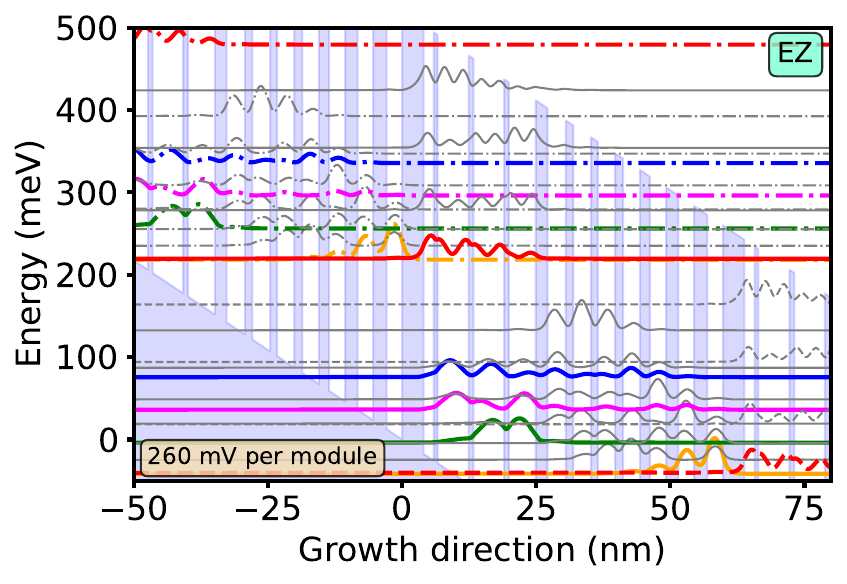}
\caption{\label{fig:BeckEZ260} EZ levels obtained from the WS levels in Fig.~\ref{fig:BeckWS260} by unitary transformations within multiplets of levels, which are close in energy (here with \texttt{gamma}$=10$ meV). See Fig.~\ref{fig:BeckWannier0} for labeling.}
\end{figure}

Based on Ref.~\onlinecite{RindertPhysRevApplied2022} this can be achieved by the following procedure:
WS-levels that are close in energy are sorted into a multiplet. The maximal energy distance is set as \texttt{gamma} in \texttt{resource\_QCL.calcEZ} with a default value of 5 meV.  Within each multiplet the matrix of the Z-operator $\hat{z}$ is calculated and diagonalized. This provides a unitary transformation for the states within the multiplet. As the levels in the multiplet have approximately the same energy, the resulting diagonal elements of the Hamiltonian are rather similar, so that the energy $E$ of the levels are reasonably well defined, suggesting the denomination as EZ levels.   
The resulting levels of this procedure are shown in Fig.~\ref{fig:BeckEZ260} for the device of Ref.~\onlinecite{BeckScience2002}. This demonstrates the spatial splitting of the  extended states close to degeneracies.

\subsection{Usage for device simulations}
In order to analyze the QCL under operation, one needs to solve a kinetic equation for the carrier distribution. This takes into account both the Hamiltonian $\hat{H}^F$ of the ideal structure and scattering processes due to deviations from ideality such as electron-phonon interaction, interface roughness, etc, see Ref.~\onlinecite{JirauschekApplPhysRev2014} for details. In principle, this kinetics can be formulated in any basis and the non-equilibrium Green's function (NEGF) technique satisfies basis invariance (albeit this is limited by truncation in the basis set and further approximations in a practical implementation). 
The NEGF code from our group \cite{WingeJAP2016} uses the basis of Wannier functions in order to cover the low-bias region. The package \texttt{resource\_QCL} described here provides the necessary input with the Hamiltonian  
and wave functions to calculate the scattering matrix elements for a specific sample.

However, NEGF codes are mumerically demanding and it is often desirable to have simple concepts at hand where the carriers can be assumed to jump between distinct levels. This is the realm of Pauli master equations. 
These had been first formulated for QCLs in the energy eigenbasis of $\hat{H}^F$, where all transition rates are due to scattering quantified by Fermi's golden rule\cite{DonovanJAP2001,IottiPRL2001}. For such approaches, the WS levels from  \texttt{resource\_QCL} can be directly applied to calculate the scattering matrix elements. As nicely explained in Ref.~\onlinecite{CallebautJAP2005} the energy eigenbasis is problematic to describe tunnel injection. A common remedy is to consider states localized on different sides of the tunnel barrier with a tunneling rate in between, which is obtained from a density matrix calculation \cite{TerazziNJP2010,JirauschekJAP2017,ChenJAP2024}. 
While commonly module eigenstates are used for this purpose, the EZ levels from \texttt{resource\_QCL} are a preferable alternative due to their orthogonality. In particular, the tunnel-matrix element can be directly extracted from the respective nondiagonal matrix element of the Hamiltonian in this basis. They are contained in \texttt{EZ.H0} and \texttt{EZ.H1} for intra- and inter-module processes, respectively, as demonstrated in \texttt{example\_to\_use\_resource\_QCL.py} from the supplementary material. Correspondingly, the matrix elements \texttt{EZ.Z0} and \texttt{EZ.Z1} for the $\hat{z}$-operator are available to quantify the optical transition strengths. 
A detailed discussion and further references regarding these different approaches can be found in Ref.~\onlinecite{WackerBook2023}.

\section{Conclusion}
We presented a concise way to establish orthonormal sets of states for QCLs which satisfy the periodicity criterion. The presented Python package \texttt{resource\_QCL} provides subroutines to establish Wannier, WS and, EZ states via unitary transformations from the initial Bloch states of the heterostructure without electrical potential. 
Each type of states is attributed a specific class
which contains the two-component wave functions, the matrices for the Hamiltonian and the $\hat{z}$-operator in the respective basis, as well as further quantities of interest.
If the mean-field $\Phi^\textrm{MF}(z)$ is known, it can be included to calculate WS and EZ states. 

The software has been tested for a large variety of QCL structures operating either in the infrared or THz range. The included plotting routine can be used to visualize the levels for a given structure. Equally important, the strictly orthonormal basis states establish a firm ground to calculate scattering matrix elements, optical transition strengths, etc, as required for quantum kinetic calculations.
 
\section*{Supplementary Material}
The supplementary material contains the Python package \texttt{resource\_QCL.py} with the codes described above. In addition, two scripts using this package are included: \texttt{beckSience2002.py} producing the figures of the article and  
\texttt{example\_to\_use\_resource\_QCL.py} demonstrating the variety of possible usage.

\section*{Acknowledgment}
The authors thank NanoLund for financial support.


%

\end{document}